\documentclass[journal=jacsat,manuscript=article]{achemso}

\usepackage[version=3]{mhchem} 
\usepackage[T1]{fontenc}       
\usepackage{color}
\usepackage{ulem} 

\author{Quanbo Jiang}
\author{Aline Pham}
\author{Martin Berthel}
\author{Serge Huant}
\affiliation{Universit\'e Grenoble Alpes, Institut NEEL, F-38000 Grenoble, France
and CNRS, Institut NEEL, F-38042 Grenoble, France}
\author{Joel Bellessa}
\affiliation{Institut Lumi\`ere Mati\`ere, UMR5306 Universit\'e Lyon 1-CNRS, Universit\'e de Lyon, 69622 Villeurbanne cedex, France}
\author{Cyriaque Genet}
\affiliation{ISIS, UMR 7006, CNRS-Universit\'e de Strasbourg, 8, all\'ee Monge, 67000 Strasbourg, France}
\author{Aur\'elien Drezet}
\email{aurelien.drezet@neel.cnrs.fr}
\affiliation{Universit\'e Grenoble Alpes, Institut NEEL, F-38000 Grenoble, France
and CNRS, Institut NEEL, F-38042 Grenoble, France}
\title[An \textsf{achemso} demo]
  {Directional and singular surface plasmon generation in chiral and achiral nanostructures demonstrated by Leakage Radiation Microscopy}

\abbreviations{LRM,SPP}

\begin{document}

\begin{tocentry}

\includegraphics[width=8.89cm]{TOC}

\section*{For Table of Contents Use Only}

Directional and singular surface plasmon generation in chiral and achiral nanostructures demonstrated by Leakage Radiation Microscopy.\\

Quanbo Jiang, Aline Pham, Martin Berthel, Serge Huant, Joel Bellessa, Cyriaque Genet, Aur\'elien Drezet.\\

The present Table of Content (TOC) Graphic illustrates the experimental setup implemented in the letter, that is leakage radiation microscopy (LRM), in order to study the directionality of propagating surface plasmon polaritons induced in achiral and chiral plasmonic structures under circular polarization excitation ($\sigma$ refering to the handedness). Specifically, the TOC Graphic depicts some LRM experimental results obtained for a ring made of elementary T-shaped apertures milled in a thin metallic film, showing both radial directionality and plasmonic vortices according to the spin excitation, with a bright (respectively dark) central spot under $\sigma=+1$ (resp.$\sigma=-1$).

\end{tocentry}

\begin{abstract}
In this paper, we describe the implementation of leakage radiation microscopy (LRM) to probe the chirality of plasmonic nanostructures. We demonstrate experimentally spin-driven directional coupling as well as vortex generation of surface plasmon polaritons (SPPs) by nanostructures built with T-shaped and $\Lambda$- shaped apertures. Using this far-field method, quantitative inspections, including directivity and extinction ratio measurements, are achieved via polarization analysis in both image and Fourier planes. To support our experimental findings, we develop an analytical model based on a multidipolar representation of $\Lambda$- and T-shaped aperture plasmonic coupler allowing a theoretical explanation of both directionality and singular SPP formation. Furthermore, the roles of symmetry breaking and phases are emphasized in this work. This quantitative characterization of spin-orbit interactions paves the way for developing new directional couplers for subwavelength routing.\\
\textbf{KEYWORDS:} Plasmonics, leakage radiation microscopy, optical chirality, optical vortex, surface plasmon directionality
\end{abstract}

Chiral plasmonic nanostructures~\cite{Chiral} exhibit unique optical properties, such as asymmetric optical transmission~\cite{Zheludev} and singular optical signatures, like vortices, visible in both the optical near-field~\cite{Yuri1,Yuri2} and in the far-field of twisted structures~\cite{Drezet1,Yuri3}.  Motivated by fundamental questions as well as by their potentials ranging from highly integrated photonic circuits to quantum optics~\cite{Chiral,revue}, interest in the field of chiral plasmonics has subsequently become a topic of intensive research. These peculiar optical effects stem from the spin-orbit interactions of light via plasmonic nanostructures, in which the photon spin couples to its spatial motion~\cite{Yuri1} and in particular to its orbital angular momentum. This leads to optical spin Hall effects~\cite{SpinHall1,SpinHall2}, i.e., to a polarization-dependent photon shift evidenced with SPPs~\cite{Yuri1,Genet2012,Spin} which can be used for instance for inducing SPP directional coupling~\cite{nif,naf,nouf}. In this context, recent studies done in the optical near-field demonstrated that spin-controlled SPP directionality and vortex generation can be achieved with chiral nanostructures such as T-shaped aperture arrays, rings or spirals milled in metal films~\cite{Tshaped,Spektor}. While the additional degree of freedom added by the incident spin enables tunable directionality, enhanced directional coupling is achieved by the broken symmetry of the plasmonic structures. Moreover, since controlling SPP propagation direction and rotational motion is essential for applications in integrated optics and optical trapping, it becomes urgent to develop sensitive imaging techniques to map plasmonic chirality not only in the near-field but also in the far-field. Due to the inherently confined SPP fields, near-field optical detection has been widely employed in the past to image SPP propagation (for a review see \cite{Drezetrevue2007}). Indirect imaging via scattering of SPPs or via grating have also been used~\cite{Grating,Subwave,Today}. Here, we propose a different approach based on leakage radiation microscopy (LRM)~\cite{Drezetrevue2008,Hecht,Drezet2013}. As a complementary method for direct imaging of SPP propagation, LRM is a powerful tool allowing quantitative analysis which has already been successfully applied to the analysis of several  in plane devices such as Bragg mirrors~\cite{DrezetOL2007}, cavities, and plasmonic crystals~\cite{DrezetNano2007,Drezetrevue2008}. Recent works also applied LRM for analyzing directional excitation of SPPs~\cite{alpha,beta}. In addition, being a far-field method, LRM is a versatile method enabling polarization analysis in both direct and Fourier spaces~\cite{Wang}. In this context, LRM has recently been applied to weak measurements of light chirality  with a plasmonic slit~\cite{Genet2012}. This pioneer work (see also refs.~\cite{naf,nouf}) showed the potential of LRM for polarization tomography and control which strongly motivates the present study of SPP directional propagation and vortex generation.\\ 
\indent In this letter, we report the implementation of LRM for probing the chirality of plasmonic nanostructures. Specifically, we achieved quantitative characterization of spin-controlled directional couplers consisting of rectangular nano-apertures arranged in T or $\Lambda$ geometry~\cite{Vshaped,Vshapedbis} (see Figure1).  On the one hand, we demonstrate unidirectional propagation of SPPs induced by a rectangular array made of the aforementioned chiral and achiral apertures. The recorded images in the Fourier space leads to direct determination of the ratio of SPP intensities along two opposite directions of propagation and defined as the directivity. Thereby, quantitative comparison between the directional coupling efficiency induced by a T-shaped aperture array and that of a $\Lambda$-shaped aperture-based array can be achieved. One important finding, which confirms previous works \cite{Tshaped,Vshapedbis}, is that both $T$- and $\Lambda$-shaped slit arrays induce SPP directionality resulting from the spin-orbit coupling mechanism. Our work provides therefore a unified perspective for $\Lambda$- and T-shaped systems by clarifying the role of phase and symmetry. Furthermore, to justify this result we introduce an analytical theory which models the radiated field by each individual rectangular aperture by a pair of electric dipoles~\cite{Yuri3}. Our analytical approach provides physical insights of the mechanism into play without requiring numerical solutions. 
On the other hand, we show that LRM is well adapted for the study of singular SPPs decoupled from the metal surface and radiated into the far-field. Using circular structures with an intrinsic handedness, (i.e. using T-shaped apertures) we demonstrate radial SPP focusing and reveal spin-orbit coupling with generation of vectorial plasmonic vortices. Finally, we show that a theoretical analysis using our dipolar model allows for both qualitative and quantitative description of such SPP singular fields.\\     
\begin{figure*}[h!t]
\centering
\includegraphics[width=16cm]{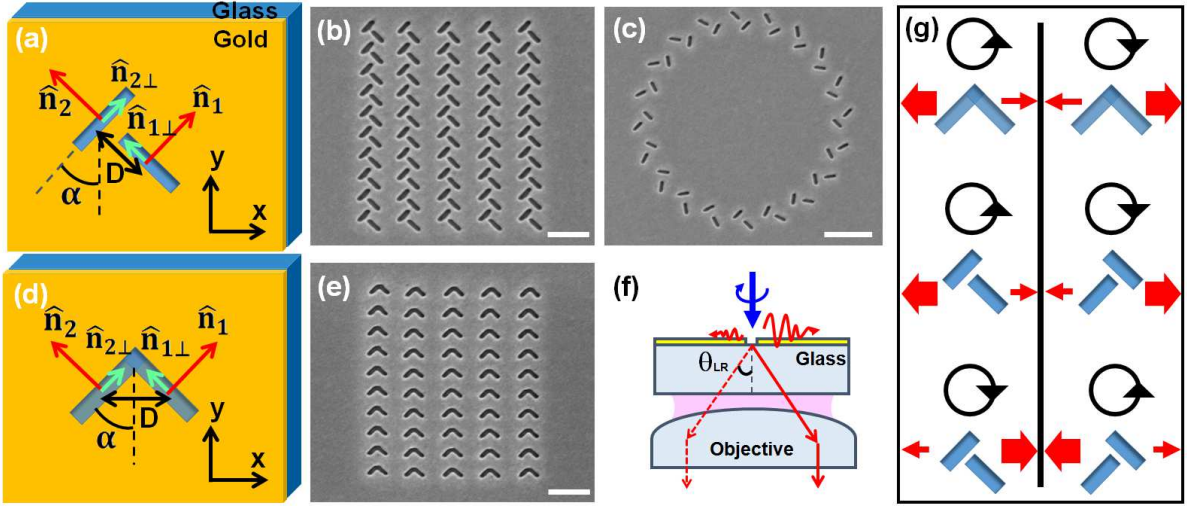}
\caption{(a, d) Schematic diagrams of a T- and $\Lambda$-shaped aperture (top view). The red arrows are the major-axis dipoles perpendicular to the slits and the green arrows are minor-axis dipoles parallel with the slits. (b), (e) SEM images of arrays made of T- and $\Lambda$-shaped apertures, respectively. (c) SEM image of 20 T-shaped apertures arranged into a ring geometry.  Scale bar value: 600 nm in (b), (e), $1 \mu m$ in (c). (f) Leakage radiation microscopy principle: The launched SPPs leak with a specific angle ($\theta_{LR}\approx44^\circ$) through the thin gold film into the glass substrate. They are then collected by an oil-immersion objective. (g) Schematic diagram illustrating the mirror symmetry based analysis for $\Lambda$- and T-shaped aperture systems illuminated by a circular polarized beam as indicated by the black arrows. The thick and thin red arrows represent the predominant and minor directions of SPP propagation, respectively.}    \label{figure1}
\end{figure*}
The directional couplers considered here are fabricated by evaporating a 50 nm thick Au layer on top of a glass substrate. The T- and $\Lambda$-like structures consist of elementary rectangular apertures, 200 nm long and 50 nm wide and arranged in T or $\Lambda$ configuration as shown in Figure \ref{figure1}(a),(d). The apertures were milled using focused ion beam (LEO 1530 30kV). The distance $D$ spacing two rectangular slits apart is fixed at 212 nm (respectively 150 nm) in case of T-shaped (resp. $\Lambda$-shaped) structures such that $k_{SPP} D/\sqrt{2}=\pi/2$ (resp. $k_{SPP} D=\pi/2$) where $k_{SPP}$ is the real part of the SPP wavevector~\cite{Subwave}. Such a phase constraint provides the necessary condition for SPP directionality as it will be explained below. The later pair of slits are arranged in 5x10 arrays with a horizontal pitch of 600 nm to achieve SPP resonant excitation in the $Ox$ direction and a vertical pitch of 300 nm for preventing SPP propagation in the $Oy$ direction (i.e., at least up to the second diffraction order).\\
\begin{figure}[h!t]
\centering
\includegraphics[width=8.5cm]{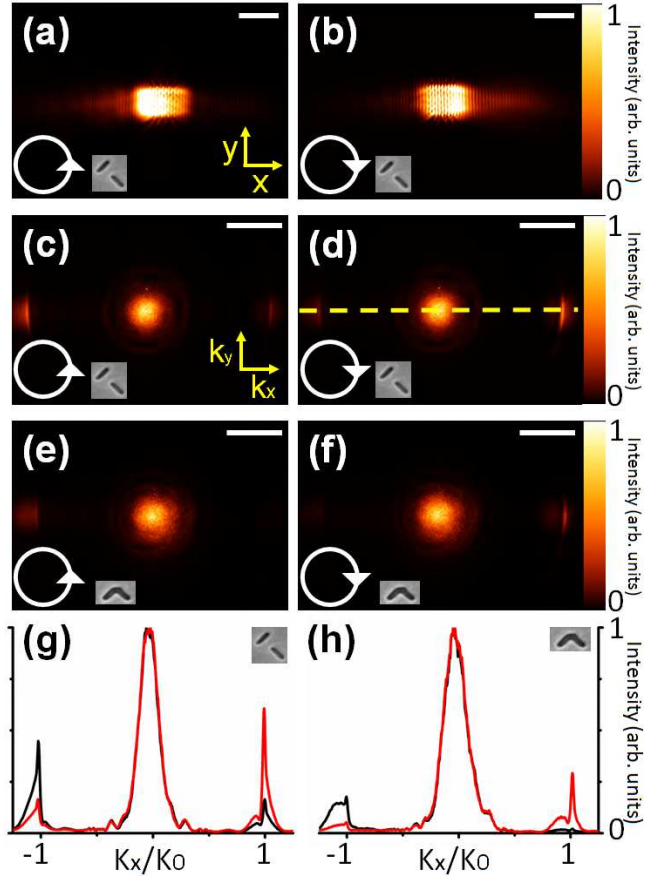}
\caption{LRM images recorded in the direct (a, b) and Fourier plane (c, d) for an array of T-shaped apertures (Figure \ref{figure1}(b)) upon LCP and RCP as indicated by the white arrows. The elements forming the arrays are displayed in SEM images in insets. (e, f) Fourier-plane images obtained for an array of $\Lambda$-shaped apertures ((Figure \ref{figure1}(e), $\alpha=60^\circ$). (g) Intensity cross-section profiles along the center line of Fourier images: the black and red curves corresponds to the cross-section in (c),(d). (h) Intensity cross-section profiles along the center line of Fourier images: the black and red curves corresponds to the cross-section in (e),(f) respectively. Scale bar value: $3\mu m$ in (a, b), $0.5k_0NA$ in (c-f).}    \label{figure2}
\end{figure}
\indent The plasmonic arrays are illuminated by a weakly focused laser at 633 nm wavelength. The incident beam is prepared either in right circular polarization (RCP) or left circular polarization (LCP) states. The excited SPPs are then recorded in the far field on a CCD camera following LRM characterization method. The LRM detection principle is sketched in Figure \ref{figure1}(f), more details on the complete setup can be found in the Supporting Information. It relies on the principle\cite{Drezetrevue2008} that in a thin metal film surrounded by two dielectric media with different refractive index, here air and glass, SPPs can radiate as leaky electromagnetic waves into the higher index medium, that is, the glass substrate. In this configuration, the wavevector matching imposes that leaky SPPs can only exit the metal film at a given angle $\theta_{LR}$ such that:
\begin{equation}
n\sin{(\theta_{LR})}=\Re {\sqrt{\dfrac{\varepsilon_{metal}}{1+\varepsilon_{metal}}}}
\end{equation}
where $\Re$ stands for the real part, $n\simeq 1.5$ is the glass optical index, $\varepsilon_{metal}$ the metal dielectric permittivity. Providing $\theta_{LR}\simeq 44\char23$is larger than the critical angle for an air-glass interface $\theta_c=42\char23$, an oil immersion objective (NA=1.45) is used below the sample to collect and transmit the leaky radiation to far-field. Thereby, imaging of SPPs in both direct and Fourier planes can be achieved. SPPs being emitted at a fixed angle $\theta_{LR}$, their intensity distribution forms a localized and bright intensity ring pattern with a radius of $k_0n\sin{(\theta_{LR})}=k_{SPP}$ in Fourier space and to which SPP polarization is orthogonal (See supporting information). The intensity along this ring is uniform in case of isotropic propagation whereas it exhibits intensity variations if SPPs propagate in selected directions. Subsequently, imaging in the reciprocal space is particularly relevant for direct mapping of spin-dependent SPP directionality where quantitative analysis can be carried out as we now demonstrate experimentally.\\ 
\indent We present in Figure \ref{figure2}(a), (b) LRM images recorded in the direct space corresponding to the excitation of the array of T-shaped apertures by LCP and RCP incident beams as indicated by the white arrow. No polarization analysis is performed at this stage. The coupler clearly exhibits spin-dependent SPP directionality: we observe a bright wake corresponding to leaky SPPs which either propagate to the left upon LCP or to the right direction upon RCP excitation. Note the fringes are due to interferences between the leakage radiation and the incident beam directly transmitted through the sample. In order to characterize the directional coupling efficiency of our plasmonic system, we introduce a quantitative criterion such as the directivity contrast which defines the capacity of the system to couple an incident spin into SPPs propagating in a given direction. It is defined as:
\begin{equation}
V_{\sigma}=\frac{\vert{I_{\sigma,+}}-I_{\sigma,-}\vert}{I_{\sigma,+}+I_{\sigma,-}}
 \label{eqn:visibility1}
\end{equation}
with $I_{\sigma,d}$ the SPP intensity launched upon $\sigma=-1$ (RCP) or $\sigma=+1$ (LCP) incident spin and propagating towards $d=+$ right or $d=-$ left direction. Owing to the widespread intensity span which decreases exponentially with the distance, the determination of the directivity in the image plane cannot be clearly defined. On the contrary, as mentioned earlier, the SPP intensity propagation towards the given direction $\pm\hat{k}_x$ translates into keen intensity peaks in the reciprocal space as depicted in Figure \ref{figure2}(c) and (d). As explained in the following,  this point allows a precise  experimental determination of the directivity given by eq.\ref{eqn:visibility1}. Note that the central bright spot originates from the incident light directly transmitted through the sample. In Figure \ref{figure2}(g), we provide intensity cross-sections performed along the center line (yellow dashed line) of images Figure \ref{figure2}(c-d). We retrieve the aforementioned central lobe as well as the two side peaks corresponding to propagating SPPs. When the array is excited by $\sigma=+1$ (black curve), one clearly measures an intense SPP peak towards the negative wavevectors $-k_x$  while it appears very weak towards $+k_x$ ($I_{+1,+}\gg I_{+1,-}$). Inversely, upon $\sigma=-1$ (red curve), propagation towards $-k_x$ exhibits an attenuated intensity peak compared to that of $+k_x$ ($I_{-1,+}\ll I_{-1,-}$). Using eq.\ref{eqn:visibility1}, the intensity contrast provided by Fourier imaging allows for determining a mean value for the directivity of $\left\langle V \right\rangle =0.59\pm0.09$ associated with the array of T-shaped apertures. We point out that the directivity measured for $\sigma=-1$ differs only slightly from that of $\sigma=+1$. Furthermore, implementing the same LRM experiment on an array made of symmetrical $\Lambda$-shaped apertures (Figure \ref{figure1}(e)) shows the later also demonstrates spin-controlled propagation of SPPs (Figure \ref{figure2}(e, f)) with predominant propagation towards $-k_x$ and $+k_x$ upon $\sigma=+1$ and $\sigma=+1$ respectively. From the intensity cross section profiles (Figure \ref{figure2}(h)), we find the array of $\Lambda$-shaped apertures to exhibit a better directional coupling efficiency than the array of T-shaped apertures with a directivity of $\left\langle V \right\rangle=0.74\pm0.07$. We thus have shown that both plasmonic devices exhibit a clear photonic spin-control of the SPP directivity along a metal film. Furthermore, LRM is shown to provide a direct method for mapping this SPP directivity. This method becomes quantitative in the Fourier plane where the contrast between the SPP peaks for propagation along $\pm k_{x}$ directions can be precisely recorded (we point out that working in the Fourier space is here mainly a practical advantage since it is also possible to work directly in the x-y plane~\cite{Tshaped,Drezetrevue2008}).\\ 
\indent In order to understand the physics involved in the directional propagation of SPPs, we now propose an analytical model based on a two-dipole representation of the nanostructures (such a model was already considered in ref.~\cite{Tshaped}). Let us first start by considering a single rectangular aperture on a thin metallic film. Upon illumination, the component of the incident beam that is mainly polarized perpendicularly to the aperture's edges is selectively scattered and coupled into SPPs, which then propagate along the gold layer perpendicularly to the aperture. A rectangular nanoaperture can thus be assumed to act as a weighted sum of two SPP in-plane dipoles oriented orthogonally: one dipole perpendicular to the short-axis W (called the minor-axis dipole) and a second dipole whose direction is perpendicular to the long axis L (major-axis dipole). The electric dipole moment associated with each rectangular aperture (labeled by $ m=1, 2$) forming a T- or $\Lambda$-shaped structure is given by: 
\begin{equation}
\overrightarrow{\mu}_{m,\sigma}=\eta[(\overrightarrow{E}_{\sigma}\cdot\widehat{n}_{m})\widehat{n}_{m}+\beta(\overrightarrow{E}_{\sigma}\cdot\widehat{n}_{m\perp})\widehat{n}_{m\perp}] \label{eqn:dipole}
\end{equation}
where $\eta$ stands for the polarizability of the major-axis dipole which depends on the geometry of the slit, $\overrightarrow{E}_\sigma $ denotes the exciting electric field of spin $\sigma$, $\widehat{n}_m$ and $\widehat{n}_{m\perp}$ indicates the major-axis and the minor-axis dipole orientation, respectively. We also introduce a coefficient $\beta$  in order to account for the weighted contribution of the minor-axis dipole with respect to the major-axis dipole in the SPP launching (the case $\beta=0$ corresponds to an infinitely long slit). In order to describe LRM radiation by these dipolar elements, we then give an intuitive analytical theory which captures all the main physical properties of SPP decoupling through a thin film in the Fourier space. Indeed, as it has been shown more rigorously elsewhere~\cite{Drezet2013,Berthel2015}, the SPP electric field signature of a single dipole $\overrightarrow{\mu}$ radiating light in the Fourier plane of a high numerical microscope objective is simply given by $\overrightarrow{E}^{SPP}_{\sigma}(\overrightarrow{k})=\widehat{k}(\overrightarrow{\mu}\cdot\widehat{k})F_{SPP}(k)$. Here, $\overrightarrow{k}$ denotes the in-plane wavevector of the radiated light. $F_{SPP}(k)\simeq 1/(k-k_{SPP})$ defines the distribution of
SPP radiated waves detected in the Fourier plane which has a Lorentzian lineshape~\cite{Drezetrevue2008}, depicting the narrow ring like shape peaked on the SPP wavevector $k_{SPP}$. The scalar product $\overrightarrow{\mu}\cdot\widehat{k}$ reveals the dipolar nature of the SPP emission and finally the unit vector $\widehat{k}$ reminds the essential longitudinal nature of the SPP electric field along the interface. The analytical model takes explicitly into account the high numerical aperture objective and the non paraxial propagation of leaky waves through the microscope~\cite{Drezet2013,Berthel2015}. Using this approach, the SPP field coherently radiated by a system of T-shaped or $\Lambda$-shaped slits is expressed by:
 \begin{equation}
\overrightarrow{E}^{SPP,}_{\sigma}(\overrightarrow{k})=\sum_{m}\widehat{k}(\overrightarrow{\mu}_{m,\sigma}\cdot\widehat{k})e^{i\phi_m}F_{SPP}(k) \label{eqn:SPP}
\end{equation}
where  $m=1,2..., N$ now label  the $2N$ dipoles associated with the $N$ individual cell constituting the arrays. The spatial phase factor $\phi_m=-\overrightarrow{k}\cdot\overrightarrow{x}_m$ accounts for the different dipole positions $\overrightarrow{x}_m$ (taken from the slit center). These phases provide the key ingredient for the coherent SPP induced spin-control studied in this work. We emphasize that our approach generalizes works done with $\Lambda$-\cite{Vshaped,Vshapedbis} and  T-shaped~\cite{Tshaped,Spektor} apertures on opaque thick film. The present approach includes a description of SPP propagation in both the direct and reciprocal planes and reduces to previous results in the limiting case $\beta=0$. Moreover, to obtain directional scattering, constructive and destructive interferences hold if $k_{SPP} D/\sqrt{2}=\pi/2$ (resp. $k_{SPP}D=\pi/2$) for T-shaped~\cite{Tshaped,Spektor} (resp. $\Lambda$-shaped) apertures. In the case of a single  aperture eq.\ref{eqn:SPP} then reads as:
\begin{equation}
\overrightarrow{E}^{SPP}_{\sigma,\pm}(k)=\pm\frac{\eta}{\sqrt{2}}F_{SPP}(k)cos^{2}(\alpha)C_{\sigma,\pm}\widehat{k}_{x\pm}
 \label{eqn:SPP3}
\end{equation}
where we introduce the coefficient $C_{\sigma,\pm}$:
\begin{equation}
\begin{aligned}
C_{\sigma,\pm}=1\pm{\sigma}\tan\alpha+(\sigma\tan\alpha\pm 1)i\\
+\beta[1\mp{\sigma\tan\alpha}-(\sigma\tan\alpha\mp{1})i]
\end{aligned}
\label{eqn:SPP3S2}
\end{equation}
which describes the interactions between the incident spin $\sigma$ and the nanostructure leading to unidirectional propagation of SPPs along $\widehat{k}_{x\pm}$. $\alpha$ is a geometrical angle defined graphically on Figure 1 (a) and (d). A precise knowledge of $C_{\sigma,\pm}$ is crucial as it determines the directional coupling efficiency of the system for a given incident spin and propagation direction. In case of a T-shaped aperture ($\alpha=45\char23$) and if we assume infinitely thin rectangular apertures, the minor dipole contribution can be neglected ($\beta$=0) and eq.\ref{eqn:SPP3S2} thus writes as:
\begin{equation}
\begin{aligned}
C_{\sigma,\mp}= 1\pm\sigma +(\sigma\pm1)i
\end{aligned}
\label{eqn:SPP2}
\end{equation}
We clearly find for a given incident spin, e.g. $\sigma=+1$, that the coupling coefficient takes its maximum value along the $-k_{x}$ direction while it is minimum towards the opposite direction $+k_{x}$ (inversely upon $\sigma=-1$). We emphasize that eq.~\ref{eqn:SPP3} is slightly modified if we consider not a single aperture but the full array. Mainly, we have to include an additional form factor  which modifies the 2D radiation pattern in the Fourier plane. However, this does not impact the main reasoning since we focus our attention on propagation in the $\pm \widehat{k}_x$ direction. Furthermore, we point out that the previous model  can be generalized to include the full expansion of transverse electric (TE) an magnetic (TM) modes radiated by the dipoles \cite{Drezet2013,Berthel2015}. This method is systematically used in the present work. For rectangular arrays it leads to results identical to the intuitive model (as far as the TE components have a weak contribution as it is the case for not too thin metal film\cite{Drezetrevue2008}).   \\  
\indent Moreover, via a careful choice of  $D$ and $ \alpha $, we can design a $\Lambda$-shaped aperture so that it can be described by the same $ C_{\sigma,\pm} $ as a T-shaped aperture. Although one features symmetrical geometry while the other is chiral, they exhibit similar spin-dependent directional coupling.   We emphasize that intuitively the array of $\Lambda$-shaped aperture is expected to induce no directional SPP coupling  along the $\pm k_x$ directions since the structure preserves the mirror symmetry with respect to the $Oy$ axis. This is different from the strategies used in refs.~\cite{nif,naf,nouf} where an oblique incident light on a non-chiral slit, by breaking mirror symmetry with respect to the $Ox$ axis, generates directional SPP coupling along the $x$ direction (with an additional   propagation along the $y$ direction). Our result follows the methods discussed in refs.~\cite{Vshaped,Vshapedbis} (see also refs.~\cite{Yannick,Grosjean,fortunobis} for related works). Here the directionality comes from a difference of sign in the coupling of light polarized along $Ox$ to SPP propagating in the $\pm k_x$ directions. This naturally respects the dipolar nature of the SPP fields. No such difference, i.e. for SPP propagating in the $\pm k_x$ directions,  occurs for a light polarization along $Oy$ in agreement with symmetries. For a circularly polarized light we have a combination of both and by choosing  the phases $\phi_m$ in eq. 4 we can obtain eqs. 6 and 7 which imply the observed SPP directionality modulated by a spin-orbit coupling.

From this important result, we can predict the behavior of an array of T-shaped aperture obtained by mirror symmetry (with respect to $Oy$ axis) of that of depicted in Figure \ref{figure1}(a). The principle is illustrated in Figure \ref{figure1}(g). Intuitively, from a mirror symmetry-based reasoning, if we know that the system in Figure \ref{figure1}(a) shows left directionality upon LCP excitation, hence we can deduce that the mirror symmetrical arrays should exhibit right directionality upon RCP excitation. However, from this reasoning, no conclusion can be drawn on the response of the system in Figure \ref{figure1}(a) upon RCP illumination. Nevertheless, from our multi-dipoles model, we have shown that the later behaves as an array of $\Lambda$-shaped apertures. Therefore we expect SPP propagation to the right upon RCP excitation. Accordingly, its mirror image should induce left propagation upon LCP illumination. This has been confirmed experimentally and the results can be found in the supporting information. In addition of revealing the similarity between $\Lambda$- and T- shaped apertures, our theory enables us to predict the SPP directional coupling of chiral plasmonic structure that could not be deduced from a pure mirror symmetrical reasoning. Remarkably, the array considered in Figure 1 (b) involves closely packed T-shaped apertures such as the distance between to Ts equals merely the distance between two slits in a T. Therefore, in each column of the array, there is approximately an equal number of T aperture and its mirror image. Consequently, for an infinite array in the $Oy$ direction the geometrical chirality of the whole structure vanishes. If this geometrical chirality of the array was a key issue here then apparently the directionality observed should depend only on the first and last rows of the array and this would make the effect very phase sensitive (e.g., to small displacements of the excitation beam). Actually, the system is very robust meaning that the main issue is not the handedness of the T-shaped apertures. Actually, since our theory provides an equivalence between T-(chiral) and $\Lambda$-(achiral) shaped apertures (i.e. for propagation in the $\pm k_x$ directions) there is no paradox. The symmetry-break necessary to induce the SPP directionality is carried by the spin-orbit coupling with the dipoles of the slits (see eq.6). The interaction between the polarized light and the two slits of a $\Lambda$- or T-shaped aperture generates a phase difference which breaks the symmetry with respect to the $Oy$ axis. This is in agreement with the dipolar nature of the SPP source which respects the symmetry carried by the excitation. Of course, the equivalence between T- and $\Lambda$- shaped  apertures is not complete and we can find differences if we  consider SPP propagation along different directions in the Fourier plane, i.e., the $k_y$ axis (we provide additional simulations confirming this point in the supporting information file).                    

Finally, inserting eq.\ref{eqn:SPP3} in eq.\ref{eqn:visibility1} leads to: 
\begin{equation}
V_{\sigma}=\frac{2\beta(1-\beta)\tan^3\alpha+2(1-\beta)\tan\alpha}{\beta^2\tan^4\alpha+(1+\beta^2)\tan^2\alpha+1}
 \label{eqn:visibilitybeta}
\end{equation}
Note the strong dependence of the directivity on the minor-axis dipole contribution embodied by $\beta$. In particular, for a given angle $\alpha$ the directivity becomes maximum if $\beta$ vanishes, i.e. if there is no effect of the minor-axis dipole, assumption usually considered~\cite{Tshaped}. However, in practice, the directivity rarely reaches 1 so the contribution of $\beta$ cannot be neglected. It is thus taken into account here.

Thanks to our experimental approach enabling direct and precise directivity contrast measurements, we can determine $\beta$ of our system. In order to measure and optimize its determination, we fabricated several $\Lambda$-shaped structures with different angles, and measured experimentally their directivity to deduce $\beta$. We find $\beta=0.51\pm0.07$ which reveals a significant contribution of the minor-axis dipole in our system. More details on the method to determine $\beta$ can be found in the supporting information.

Noteworthy, our theory reveals that maximum directivity is reached for $\alpha=60\char23$. This important result is consistent with previous numerical simulations~\cite{Vshaped} which also predict maximum directional coupling for $\Lambda$-shaped structure array with $\alpha=60\char23$. Although a finite difference time domain~\cite{Vshaped} based numerical resolution approach was adopted (see also \cite{Yannick} for related studies with lithographed $\Lambda$-shaped antennas but with a directionality along the $ky$ axis), we showed that our analytical model based on dipolar approximation is in complete agreement with these results. Accordingly, eq.\ref{eqn:visibilitybeta} points out the dependence of the directivity on the angle $\alpha$ which explains why we have experimentally measured a better unidirectional coupling effect for the $\Lambda$-shaped apertures with $\alpha=60^\circ$ than for the T-shaped apertures.

\indent We now extend the above capability and methodology to radial directional coupling by spatially rotating the T-shaped apertures following a circle. Radial SPP propagation with inward or outward direction according to the handedness of the excitation beam is expected from near-field measurements~\cite{Tshaped}. By means of LRM, we performed detailed characterization of SPP propagation and intensity distribution with polarization analysis in both image and Fourier spaces. The structure is demonstrated, both analytically and experimentally, to induce singular SPP modes and spin-dependent radial directionality as a result of optical spin-orbit interaction.\\ 
\begin{figure*}[h!t]
\centering
\includegraphics[width=16cm]{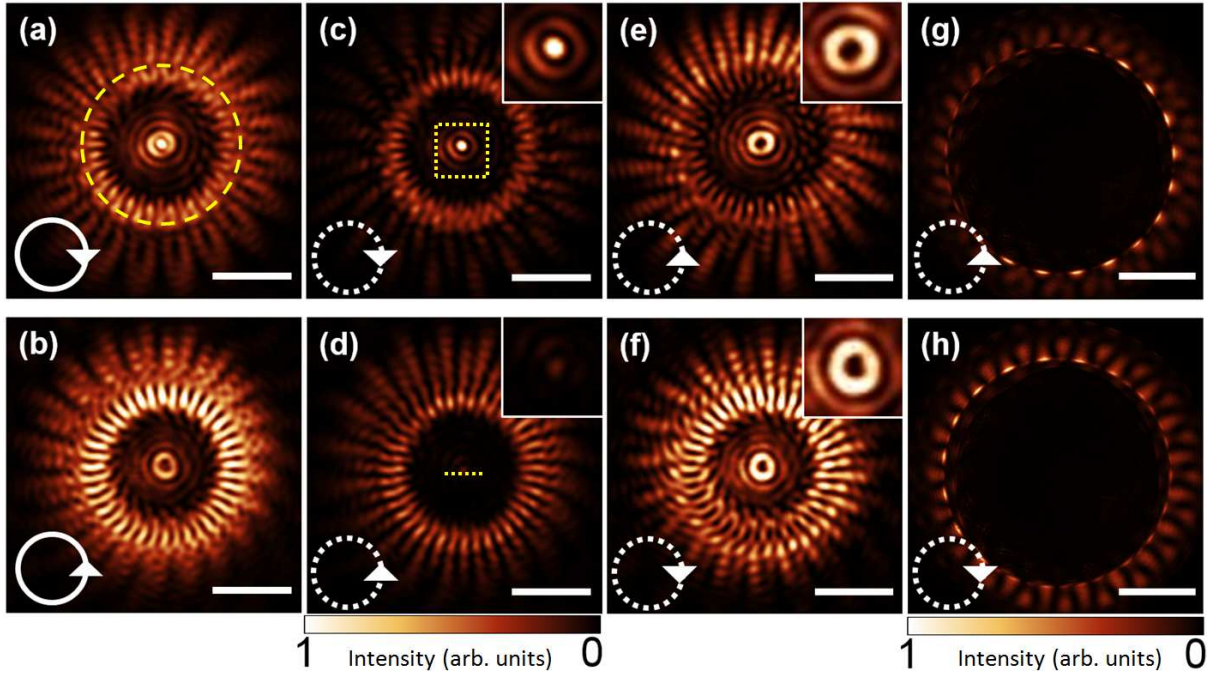}
\caption{\label{Ring T Experiments} Experimental LRM results for the plasmonic structure depicted in Figure \ref{figure1}(c). It is illuminated either with RCP (first row) or LCP (second row) polarization as indicated by the white solid arrows in (a) and (b). The yellow dashed circle in (a) indicates the position of the structure. (a-f) Transmitted LRM signals measured in the image plane (scale bar value: 2$\mu$m) and (g-h) in the Fourier space (scale bar value $0.5k_{0}NA$). (a-b) Signal recorded without and (c-h) with polarization analysis in the circular basis as indicated by the dotted arrows. Zoom on SPP focusing spot and SPP vortex are shown in insets. Exposure time 2s (a-d) and 4s (e-f). A beam block in the Fourier plane is used to remove the directly  transmitted light from the incident beam.}
\end{figure*}
\indent We fabricated a plasmonic structure where the elementary T-shaped apertures are spaced 600 nm apart, oriented and arranged into a ring of 2$\mu$m radius. The broken symmetry of the T-shaped elements imparts chirality to the structure geometry. However, we emphasize that the key feature here is the handedness associated with the whole structure and not the individual chirality of the T-elements. The plasmonic structure under study is globally right-handed as displayed in the SEM image in Figure 1 (c). A laser beam prepared in circular polarization state is weakly focused onto the sample and couples into SPPs which are then imaged on a CCD camera by LRM. The LRM signals are shown in Figure \ref{Ring T Experiments} and are collected in both image and Fourier spaces thus providing the spatial and angular distributions of the SPP intensity. A beam block is used in order to suppress the incident light directly transmitted through the sample. 
The results depicted in the first (respectively second) row of Figure \ref{Ring T Experiments} are recorded with an exciting beam prepared in RCP (LCP) polarization state as indicated by the solid white arrow.

One can readily observe on Figure \ref{Ring T Experiments}(a), (b) that SPPs radially propagate toward the center of the system or outward, depending on the incident spin. Upon RCP excitation, inward directional coupling is evident as a well defined bright spot (Figure \ref{Ring T Experiments}(a)), whereas a dark central spot (Figure \ref{Ring T Experiments}(b)) corresponding to outward propagation is observed when the system is illuminated with the orthogonal LCP state. Therefore, the chirality of our system results in a spin-sensitive radial directional coupling, as anticipated by our previous study. Our chiral plasmonic device demonstrates spin-based switching capability with a tightly focused spot that can be switched on and off by an appropriate input spin-state. Furthermore working with a thin metal film allows us to decouple the singular SPP field into the far-field.

An additional observation we address here is the remarkable annular intensity patterns surrounding the central spots. Ring profiles are typical of vortex fields. Previous works on Bull's eyes and circular apertures reported that structures with rotational symmetry create Bessel beams according to spin-states \cite{Lombard2008,Osorio2015}. To further probe the recorded SPP intensity distribution, we perform polarization analysis in both direct and Fourier spaces. The images were recorded in circular polarization basis as indicated by the dashed arrows in Figure \ref{Ring T Experiments}. Upon RCP and LCP projection, Figure \ref{Ring T Experiments}(c) and (d) display maxima intensity near the center when the signal is recorded with the same handedness as the input state whereas intensity minima with annular pattern (Figure \ref{Ring T Experiments}(e),(f)) are obtained for the orthogonal output state. Furthermore, spin-orbit coupling induced by the handedness of the structure results in brighter intensity in case of RCP excitation. The spin-driven SPP intensity and the pattern which are associated with SPP vortices will be discussed below. Noteworthy, inspection of Fourier images (Figure \ref{Ring T Experiments}(g, h)) shows the two main contributions to the observed LRM signals: the radially excited SPPs and the diffraction emission. The discretized SPP pattern demonstrates that SPPs are launched by T elementary apertures. For a given $k_{SPP}$ direction in the reciprocal space, the images corresponding to RCP and LCP output states exhibit the same SPP intensities in accordance with the convers ion of energy between SPP propagating toward and outward the center of the structure. The asymmetric SPP intensity may be ascribed to a weak misalignment of the setup or to a slight ellipticity, hence providing sensitive information on experimental errors. 

To get physical insights on the observed LRM signals, we analytically derive the solutions describing the SPP intensity distribution near the origin, generated by our plasmonic structure. In the present analytical model, the system is described by the aforementioned pair of dipoles and the structure diameter is assumed to be large enough with respect to $\lambda_{SPP}$ and separation $D$ (See Suporting information). Upon RCP illumination, SPP intensity distribution near the center is given by $I_R^{tot}=I_{R,R}+I_{R,L}$ with:
\begin{equation}
\begin{split}
I_{R,R}\propto \mid C_{-1}\mid ^2 J_{0}^2(k_{SPP}\rho )\\
I_{R,L}\propto \mid C_{-1}\mid ^2 J_{2}^2(k_{SPP}\rho )\\
\end{split}
\end{equation}
and in case of LCP illumination it writes $I_L^{tot}=I_{L,L}+I_{L,R}$ with:
\begin{equation}
\begin{split}
I_{L,L}\propto \mid C_{+1}\mid ^2 J_{0}^2(k_{SPP}\rho )\\
I_{L,R}\propto \mid C_{+1}\mid ^2 J_{2}^2(k_{SPP}\rho )\\
\end{split}
\end{equation}
$I_{i,j}$ denotes the resultant SPP intensity launched by $i=$RCP, LCP polarization states and analyzed in $j=$RCP, LCP states. $\rho$ represents the distance between the T-shaped elements and the circular structure center, and $J_l$ stands for the $l^{th}$ order Bessel function. $C_{-1}$ (respectively $C_{+1}$) refers to the SPP coupling efficiency of the T-shaped element upon RCP (resp. LCP) polarization input state as described previously (see eq.~7).\\ 
\indent In accordance with our LRM measurements, our model successfully predicts the spin orbit coupling induced by our chiral structure. It demonstrates the conversion of the incident spin into singular SPP fields which can be described by $0^{th}$ and $2^{nd}$ order Bessel functions, respectively. Similarly to a ring aperture under circularly polarized field, the structure can excite SPP vortices of topological charges of 0 and 2. However,  the rotational symmetry of the structure is broken due to the chirality ascribed to the T-shaped elements (we note however that similar results can be obtained with $\Lambda$-shaped apertures forming a circle and inducing therefore a global twist to the system. This emphasizes again the role played by phase in our experiments due to the previously discussed equivalence between $\Lambda$- and T- shaped apertures. We show in the supporting information file some experimental results confirming this point). It is associated with a  selection rule leading to SPP intensity contrast between RCP and LCP excitation. This selection rule is comprised in the spin-sensitive coupling coefficient $C_{-1}$ and $C_{+1}$ with $C_{-1}\gg C_{+1}$ (see eq.~7). In agreement with our experimental results, our calculations well reproduce the spin-sensitive SPP coupling efficiency: upon RCP illumination, high-intensity SPP fields are measured (Figure \ref{Ring T Experiments} (c), (e)) whereas lower intensity SPP fields are collected upon LCP illumination (Figure \ref{Ring T Experiments} (d), (f)). 

Moreover, this field distribution upon RCP and LCP excitation are spatially separated near the origin which is reminiscent of a spin Hall effect. Generally applied to electrons and originating from the spin-orbit coupling, spin Hall effect manifests itself as a dependence of the electron's spatial trajectory with its spin. Here, we observe a spin sensitive deviation upon interaction with the chiral plasmonic system. We can envision the implementation of the later as a mean  to generate singular SPP fields with RCP or LCP states.

\begin{figure}[h!t]
\centering
\includegraphics[width=8.5cm]{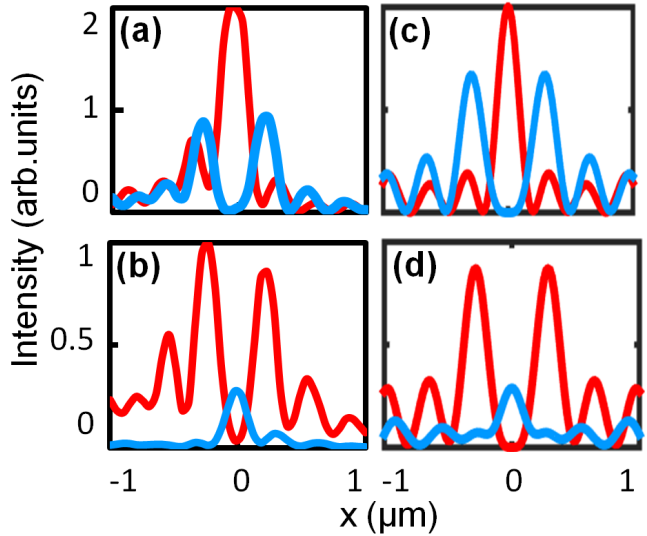}
\caption{\label{Cross-section} Intensity cross-section along center-line as indicated by the yellow dashed line in Figure \ref{Ring T Experiments}(d). (a,b) Experimental and (c, d) simulated results obtained with input states prepared in RCP (a, c) and LCP (b, d). The blue curve corresponds to output states prepared in LCP and the red curve to RCP. }
\end{figure}

To quantify the value of the SPP switching capability, we perform cross-sections along the center-line on Figure \ref{Ring T Experiments}(c-f) as indicated by the yellow dashed line in Figure \ref{Ring T Experiments}(d). The intensity cross-sections are displayed in Figure \ref{Cross-section}(a), (b). LRM images allow direct and quantitative determination of the plasmonic device features such as focusing efficiency and extinction ratio. We found a subwavelength focused spot of 247 nm FWHM in case of input and output in RCP state and an extinction ratio of 7.45 can be achieved by the chiral structure. Thus our characterization method allows quantitative measurements and can be implemented for plasmonic device performance optimization.

\begin{figure*}[h!t]
\centering
\includegraphics[width=16cm]{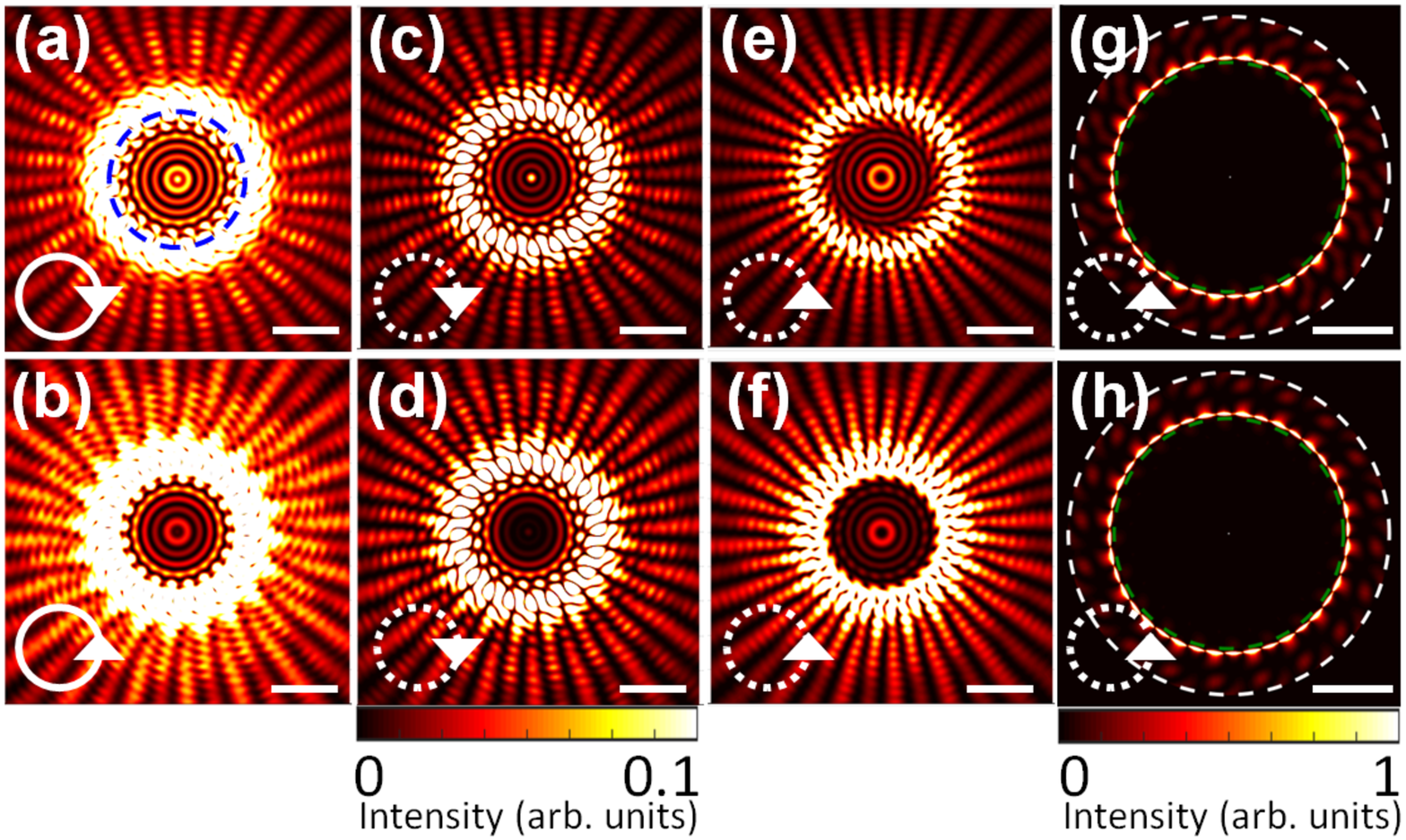}
\caption{\label{Ring T Simulations} Simulated LRM results for the structure depicted in Figure \ref{figure1}(c). It is illuminated either with RCP (first row) or LCP (second row) polarization as indicated by the white solid arrows. The blue dashed circle in (a) indicates the position of the structure (resp. beam blocker). (a-f) Transmitted LRM signals provided in the image plane (scale bar value: 2$\mu$m) and (g-h) in the Fourier space (scale bar value $0.5k_{0}NA$). The white dashed circle represents $k_{0}NA$. (a-b) Signal recorded without and (c-h) with polarization analysis in the circular basis as indicated by the dotted arrows. Exposure time 2s (a-d) and 4s (e-f). Beam block in the Fourier plane is used to block transmitted light from the incident beam.} 
\end{figure*}

The above simplistic theoretical model provided us with a physical comprehension of the mechanism into play and qualitatively described the observed SPP intensity patterns with Bessel functions. However, the contribution of diffraction and the effects of the imaging systems on the LRM images were ignored. In order to carry out quantitative analysis, we take these effects into consideration and perform simulations with the radiated fields described by TM and TE fields. We also account for the contribution of $\beta$ previously determined experimentally. It is beyond the scope of this paper to provide a full theoretical description of LRM imaging of SPPs~\cite{Drezet2013}. The simulations results are depicted in Figure \ref{Ring T Simulations}.

\indent We found the experimental and simulated results to be in excellent agreement. The simulation clearly exhibits  the fact that the right handed plasmonic structure radially focuses the incident RCP into a central peak and LCP excitation leads to a doughnut-shaped pattern. A quantitative comparison is achieved by performing cross-sections along the center-line on Figure \ref{Ring T Simulations}(c-f) similarly to Figure \ref{Ring T Experiments}(d). The intensity cross-sections are displayed in Figure \ref{Cross-section}(c), (d). We find that experimental results well support the theoretical model which predicts an extinction ratio of 6.50 and a focused spot size of 182 nm. This close inspection also reveals some differences between the measured and simulated SPP intensity profiles. Experimental uncertainties such as setup misalignment, polarizers imperfections are assumed to be at the origin of the observed discrepancies. Additionally, interactions between dipoles have been neglected in our theoretical model and hence can also contribute to those differences. These effects require further investigation such as numerical simulations accounting for inter-dipole interactions.\\
\indent To conclude, by means of  LRM, our plasmonic systems   made of T- or $\Lambda-$ shaped proved to induce spin-controlled directionality decoupled into the far-field. This is expected to play an important role for plasmonics information processing. Furthermore, comparison between simulation and experimental data showed that our theoretical model well reproduce the experimental results and allows precise determination of dipole contributions in the SPP radiated field. Simultaneously, using chiral circular gratings, radial propagation and singular SPP formation were demonstrated with LRM. By selecting the proper input and output polarization states we mapped SPP Bessel vortices in the direct space. We have shown that LRM characterization method makes quantitative analysis possible for polarization tomography therefore offering new possibilities for device development involving chirality  and surface plasmons. In particular, future work will focus on applications into the quantum domain using single photons coupled to near-field optics~\cite{Cuche2010,Jun2015,Berthel2015,Berthel2016}. All these findings are expected to have a huge impact in the field of chiral nano-plasmonics.

\begin{acknowledgement}

This work was supported by Agence Nationale de la
Recherche (ANR), France, through the SINPHONIE (ANR-12-NANO-0019) and
PLACORE (ANR-13-BS10-0007) grants. Cyriaque Genet also thanks the ANR Equipex "Union" (ANR-10-EQPX-52-01). The Ph.D.
grant of A. Pham by the Minist\`ere de l'enseignement et la recherche, scientifique, and  of Q. Jiang by the R\'egion Rh\^one-Alpes is gratefully
acknowledged. We thank J.-F. Motte and G.~Julie, from NANOFAB facility in Neel Institute, for sample fabrication.

\end{acknowledgement}

\begin{suppinfo}

Leakage radiation microscopy experimental setup. Multidipolar representation for analytical description of array and circular plasmonic systems made of T-shaped apertures. Method for determination of minor-axis dipole weight $\beta$. Experimental results measured for T-shaped aperture array and both T- and $\Lambda$-shaped aperture ring with their mirror symmetry images.

\end{suppinfo}


\end{document}